%% file: ms.tex
\documentclass[conference,compressed,twocolumn]{IEEEtran}

    \usepackage[pdftex]{graphicx}

\usepackage{color,colortbl,multicol,amssymb,amsfonts,array}
\usepackage[cmex10]{amsmath}
\usepackage{gensymb,textcomp}
\usepackage[ansinew,utf8]{inputenc}
\usepackage[acronym,shortcuts,nomain,nopostdot]{glossaries}
\usepackage{blindtext}

\definecolor{hyp}{rgb}{0,0,0.3} 
\usepackage[colorlinks=true,linkcolor=hyp,filecolor=hyp,citecolor=hyp,urlcolor=hyp]{hyperref}

\input{acronyms}

\newcolumntype{L}[1]{>{\raggedright\let\newline\\\arraybackslash\hspace{0pt}}m{#1}}
\newcolumntype{C}[1]{>{\centering\let\newline\\\arraybackslash\hspace{0pt}}m{#1}}
\newcolumntype{R}[1]{>{\raggedleft\let\newline\\\arraybackslash\hspace{0pt}}m{#1}}

\begin{document}

\author{\IEEEauthorblockN{ Stephan Jaeckel\IEEEauthorrefmark{1}, Leszek Raschkowski\IEEEauthorrefmark{1}, Frank Burkhardt\IEEEauthorrefmark{2} and Lars Thiele\IEEEauthorrefmark{1}}
\IEEEauthorblockA{\IEEEauthorrefmark{1} Fraunhofer Heinrich Hertz Institute, Berlin, Germany, stephan.jaeckel@hhi.fraunhofer.de}\IEEEauthorblockA{\IEEEauthorrefmark{2} Fraunhofer Institute for Integrated Circuits, Erlangen, Germany}}

\title{A Spatially Consistent Geometric D2D Small-Scale Fading Model for Multiple Frequencies}

\maketitle

\begin{abstract}
The 3GPP \ac{NR} channel model introduced spatial consistency and a correlation model for multiple frequencies. Future extensions of this model will incorporate mobility at both ends of the link. These features are essential for many emerging wireless technologies in the 5G era. However, the existing \ac{SSF} model does not integrate these features coherently. To solve this problem, we propose a new \ac{SSF} model that seamlessly integrates with the remaining NR model and allows the simultaneous simulation of all three features. We demonstrate this integration by showing that the output of the new SSF model agrees well with \ac{LSF} parameter distributions provided by 3GPP. This enables the simulation of new wireless technology proposals that were difficult to realize with existing \acp{GSCM}.
\end{abstract}

\section{Introduction}

\Acp{GSCM} are a well-established tool to model wireless propagation channels. They consist of two main components: a stochastic part that generates a random propagation environment, and a deterministic part that lets \acp{TX} and \acp{RX} interact with this environment. To predict the wireless system performance, the random environment must fulfill certain statistical properties which are determined by measurements. These properties are generated by the so-called \ac{LSF} model. A subsequent \ac{SSF} model generates individual \acp{MPC} for each \ac{MT}. \Acp{GSCM} became widely used by the \ac{3GPP} which required standardized models to evaluate new technology proposals. This was provided by the \ac{SCM} in 2003 \cite{3gpp_tr_25996_v610}. Since then, this model has undergone many iterations to support new features of the fast evolving wireless world. However, the \ac{SSF} model has not been significantly enhanced which leads to incompatibilities with some newly introduced features of the \ac{NR} model \cite{3gpp_tr_38901_v1500}.

The \ac{NR} model proposes to correlate all random variables that determine the powers, delays and angles of the \acp{MPC}. This so-called spatial consistency solves one major drawback of previous \acp{GSCM}, namely the lack of realistic correlation in multi-user wireless channels. An efficient way to do this is by utilizing the \ac{SOS} method in \cite{Jaeckel2018}. However, to be truly consistent, any function that modifies these random variables must be continuous. For example, the delay generation in \cite{3gpp_tr_38901_v1500} requires to sort the delays (\cite{3gpp_tr_38901_v1500}, eq. 7.5-2). Sorting is not a continuous function since it changes the order of the \acp{MPC} depending on their delay. As a result, the channel coefficients show sudden ``jumps'' when plotting the phase over time on a continuous \ac{MT} trajectory. The same happens for the scaling with the maximum path power when generating the arrival and departure angles (\cite{3gpp_tr_38901_v1500}, eqs. 7.5-9 and 7.5-14), the positive or negative sign in the angles (eq. 7.5-16), and the random coupling of rays within a multipath cluster. All these operations break the spatial consistency. The \ac{NR} model also introduces an alternative channel generation method to support multi-frequency simulations, e.g., for combined sub-6-GHz and mm-wave channels. The rationale is that, for a given environment, a \ac{MT} ``sees'' the same propagation paths (the same clusters), but with different power. Hence, delays and angles are kept fixed and path-powers are modified to account for the different \ac{KF}, \ac{DS} and \acp{AS} at different frequencies. However, the proposed method does not ensure that the output of the \ac{SSF} model (the channel coefficients) is consistent with the input (the \ac{LSF} parameters). Lastly, emerging wireless technologies increasingly require the support for \ac{D2D} communications, such as in vehicular networks, air-to-ground communications, industrial communications, or communication scenarios involving satellites in low-earth orbit. All of these use cases require that both ends of the link are mobile.

This paper addresses these issues by proposing a new \ac{SSF} model which replaces steps 5, 6, and 7 of the \ac{3GPP} \ac{NR} channel generation procedure (see \cite{3gpp_tr_38901_v1500}, pp.\ 32). The method is introduced in Section~\ref{sec:ss_model}. Numeric results are presented and discussed in Section~\ref{sec:ss_model_numeric}.  An open-source implementation is available as part of the \acF{QuaDRiGa} \cite{quadriga_www}.

\section{A New Small-Scale Fading Model}\label{sec:ss_model}

The communication model consists of multiple \acp{TX} and multiple \acp{RX}. Their locations are given in  \ac{3-D} metric Cartesian coordinates as $(x_t,y_t,z_t)$ and $(x_r,y_r,z_r)$, respectively. A transmitted signal is reflected and scattered by objects in the environment such that multiple copies of that signal are received by the \ac{RX}. Each signal \emph{path} consists of a departure direction at the \ac{TX}, a \acF{FBS}, a \ac{LBS}, and an arrival direction at the \ac{RX}. Departure and arrival directions are given in geographic coordinates consisting of an azimuth angle $\phi$ and an elevation angle $\theta$\footnote{$\phi$ is defined in mathematic sense, \emph{i.e.}, seen from above, a value of 0 points to the east and the angles increase counter-clockwise. $\theta$ is oriented relative to the horizontal plane. Positive angles point upwards.}. The formulation of the path generation procedure is done for the dual-mobility case, where both ends of the link can be mobile. For any two channel realizations, the \ac{TX} can be in a different position with $d_t$ describing the distance between the two positions. Likewise, the \ac{RX} can be displaced by a distance $d_r$. Single-mobility is a special case where either $d_t$ or $d_r$ is zero.  All random variables that determine the positions of the \ac{NLOS} scatterers are spatially correlated, \emph{i.e.}, they depend on the positions of the \ac{TX} and the \ac{RX} \cite{Jaeckel2018}.

\begin{figure}[h]
    \centering
    \includegraphics[width=84mm]{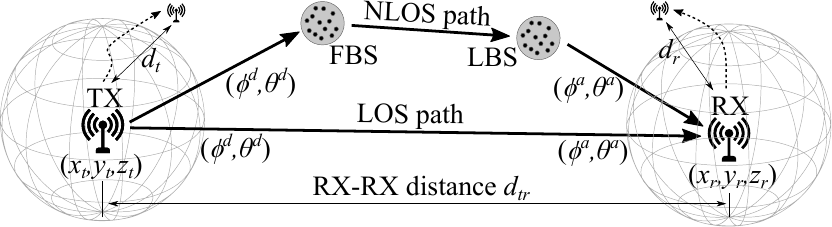}
    \vspace{-0.5\baselineskip}
    \caption{Illustration of the communication model}
    \label{fig:comm_model}
\end{figure}

\paragraph{Initial Delays and Angles}

Initial delays for the \ac{NLOS} paths are drawn randomly from a single-sided exponential distribution with unit mean and unit variance as
\begin{equation}\label{eq:initial_delays_n}
    \tilde{\tau}_l = - \ln\left\{ X^\tau_l( x_t,y_t,z_t,x_r,y_r,z_r )  \right\}\text{,}
\end{equation}
where the index $l$ denotes the path number and $X^\tau_l \sim \mathcal{U}(0,1)$ is a spatially correlated uniformly distributed random variable having values between 0 and 1. The \ac{LOS} delay, \emph{i.e.}, the delay of the first path, is set to 0. The initial delays are not scaled by the \ac{DS} nor are the angles scaled by the \ac{AS}. This approach is different compared to the \ac{NR} model \cite{3gpp_tr_38901_v1500} which includes the spreads already in the initial values. The \acF{ACF} of $X^\tau_l$ is a combination of a Gaussian and an exponential \ac{ACF}
\begin{equation}
    \rho_\tau(d) =
    \left\{
      \begin{array}{ll}
        \exp\left(-\frac{d^2}{d_\lambda^2}\right), & \hbox{for $d < d_\lambda$;} \\
        \exp\left(-\frac{d}{d_\lambda}\right), & \hbox{for $d \geq d_\lambda$,}
      \end{array}
    \right.
\end{equation}
where $d$ is the distance between two \acp{MT} positions and  $d_\lambda$ is the decorrelation distance, \emph{i.e.}, the distance at which the correlation falls below $e^{-1}\approx 0.37$ \cite{Gudmundson1991_Correlation}. The combined \ac{ACF} produces a higher correlation between delays of closely spaced \acp{MT}. It was found that this is more realistic compared to an exponential \ac{ACF} where the delays may vary significantly even when \acp{MT} move only a few centimeters. If \ac{TX} and \ac{RX} are swapped, path delays must be identical. This is known as channel reciprocity and can be modeled by generating random variables $X^\tau_l$ in \eqref{eq:initial_delays_n} as
\begin{multline}
    X^\tau_l( x_t,y_t,z_t,x_r,y_r,z_r ) = \\
    \frac{1}{2} \mathrm{erfc}\left( - \frac{ \tilde{X}^\tau_l(x_t,y_t,z_t) + \tilde{X}^\tau_l(x_r,y_r,z_r)}{2 \cdot \sqrt{\rho_\tau(d_{tr}) + 1}} \right)\text{,}
\end{multline}
where the complementary error function maps the spatially correlated normal distributed random variables $\tilde{X}^\tau_l$ to a uniform distribution required by \eqref{eq:initial_delays_n}. The scaling with $\sqrt{\rho_\tau(d_{tr}) + 1}$ ensures that the variance of the random process does not change for small \ac{TX}-\ac{RX} distances. As for the \ac{DS}, the initial values for the angles are generated spatially consistent having the same \ac{ACF} and decorrelation distance. The \ac{NR} model \cite{3gpp_tr_38901_v1500} proposes to obtain the azimuth angles from a wrapped Gaussian distribution, the elevation angles from a wrapped Laplacian distribution, and the path power from a single slope exponential \ac{PDP}. However, this leads to large angle offsets when the powers are small and the \ac{AS} is large. Due to the wrapping operation, the achievable \ac{AS} is limited. In order to have a larger range of possible angular spreads, we propose to draw all initial \ac{NLOS} angles from a uniform distribution and apply the correct \ac{AS} by a subsequent scaling operation. If \ac{TX} and \ac{RX} change places, channel reciprocity requires that the departure angles at the \ac{TX} become the arrival angles at the \ac{RX} and vice-versa. This effect is captured by generating two independent spatially correlated normal distributed random variables $\tilde{X}^{A}_l$ and $\tilde{X}^{B}_l$ and combining them to
\begin{multline}
    \tilde{\phi}_l^d( x_t,y_t,z_t,x_r,y_r,z_r ) = \\
    \frac{\pi}{2} \mathrm{erfc}\left( -\frac{ \tilde{X}^{A}_l( x_t,y_t,z_t) + \tilde{X}^{B}_l( x_r,y_r,z_r) }{2}  \right) - \frac{\pi}{2}\text{,}
\end{multline}
\vspace{-\baselineskip}
\begin{multline}
    \tilde{\phi}_l^a( x_t,y_t,z_t,x_r,y_r,z_r ) = \\
    \frac{\pi}{2} \mathrm{erfc}\left( -\frac{ \tilde{X}^{B}_l( x_t,y_t,z_t) + \tilde{X}^{A}_l( x_r,y_r,z_r) }{2}  \right) - \frac{\pi}{2}\text{,}
\end{multline}
where the complementary error function maps the resulting values to $\mathcal{U}\left(-\frac{\pi}{2},\frac{\pi}{2}\right)$. The same procedure is repeated for the initial elevation angles $\tilde{\theta}_l^d$ and $\tilde{\theta}_l^a$. The initial angles for the \ac{LOS} path are set to 0.

\paragraph{Initial Path Powers}

The initial delays $\tilde{\tau}_l$ and the initial angles $\tilde{\phi}_l^d$, $\tilde{\phi}_l^a$, $\tilde{\theta}_l^d$, and $\tilde{\theta}_l^a$ are assumed to be frequency-independent, \emph{i.e.}, a \ac{MT} sees the same scattering clusters at different frequencies and therefore, the same angles and delays are used. However, \ac{DS} and \ac{AS} are generally frequency-dependant \cite{Peter2016b}. For example, the \ac{DS} at a carrier frequency of 10~GHz is generally shorter than at 2~GHz. The \ac{NR} model \cite{3gpp_tr_38901_v1500} proposes an optional method which can adjust the path powers such that different delay and angular spreads can be achieved. The powers are calculated as
\begin{multline}\label{eq:multi_freq_cluster_powers}
    \tilde{P}_{l,f} =
        \exp\left\{ -\tilde{\tau}_l  g^\text{DS}_f \right\} \cdot
        \exp\left\{ -(\tilde{\phi}_l^d)^2  g^\text{ASD}_f \right\} \cdot \\
        \exp\left\{ -(\tilde{\phi}_l^a)^2  g^\text{ASA}_f \right\} \cdot
        \exp\left\{ -|\tilde{\theta}_l^d|  g^\text{ESD}_f \right\} \cdot
        \exp\left\{ -|\tilde{\theta}_l^a|  g^\text{ESA}_f \right\}\text{,}
\end{multline}
where the index $f$ refers to the $f=1\ldots F$ carrier frequencies. The scaling coefficients $g^\text{DS}_f$, $g^\text{ASD}_f$, $g^\text{ASA}_f$,  $g^\text{ESD}_f$, and $g^\text{ESA}_f$ need to be calculated such that the frequency-dependent differences in the spreads are reflected in the powers. For single-frequency simulations, the \ac{NR} model \cite{3gpp_tr_38901_v1500} uses the delay distribution proportionality factor $r_\tau$ to shape the \ac{PDP}. $r_\tau$ typically has values in between 1.7 and 3.8 and it follows from (\cite{3gpp_tr_38901_v1500}, eq. 7.5-5) that $g^\text{DS} = r_\tau - 1$. However, for the multi-frequency simulations, there might be a different \ac{DS} for each frequency. In this case, $g^\text{DS}_f$ is a function of the frequency and so is $r_\tau$. Hence, it is not possible to fix $r_\tau$ to a given value. Instead, we propose to calculate $g^\text{DS}_f$ by fitting a logarithmic function to the delays \eqref{eq:initial_delays_n} and powers $\tilde{P}_{l,f} = \exp\{ -\tilde{\tau}_l  g^\text{DS}_f \}$ as
\begin{equation}
    g^\text{DS}_f = -1.5 \cdot \ln\left\{ 1.2 \cdot \overline{\text{DS}}_f - 0.15 \right\}\text{.}
\end{equation}
The \ac{DS} values must be normalized to obtain a value $\overline{\text{DS}}_f$ for each frequency.
\begin{multline}
    \overline{\text{DS}}^*_f =  \frac{\text{DS}_f}{ \max_{\forall f}\left\{\text{DS}_f\right\} + \min_{\forall f}\left\{\text{DS}_f\right\}}\\
    \overline{\text{DS}}_f = \max\left\{\min\left( \overline{\text{DS}}^*_f, 0.15 \right), 0.85 \right\}
\end{multline}
If there is no frequency dependency of the \ac{DS}, then $g^\text{DS}_f$ will have a value of 1.2 which corresponds to $r_\tau = 2.2$. A similar mapping is done for the \ac{AS}. The initial angles \eqref{eq:initial_angles_90} are uniformly distributed having values in between $-\pi/2$ and $\pi/2$. As for the \ac{DS}, the scaling coefficients for the angular spreads $g^\text{ASD}_f$, $g^\text{ASA}_f$, $g^\text{ESD}_f$, and $g^\text{ESA}_f$ are obtained by
\begin{eqnarray}
  g^\text{ASD/A}_f &=& -2.2 \cdot \ln\left\{ 1.5 \cdot \overline{\text{AS}}_f - 0.35 \right\} \\
  g^\text{ESD/A}_f &=& -3.4 \cdot \ln\left\{ 1.2 \cdot \overline{\text{AS}}_f - 0.1 \right\} \\
  \overline{\text{AS}}_f &=& \min\left( \frac{ 0.75\cdot \text{AS}_f}{\max_{\forall f}\left\{\text{AS}_f\right\}}, 0.25 \right)  \text{.}
\end{eqnarray}
If there is no frequency dependency, then $g^\text{ASD/A}_f$ will have a value of 0.56 which corresponds to an initial azimuth spread of 42\degree . The scaling coefficient $g^\text{ESD/A}_f$ will have a value of 0.76 which corresponds to an initial elevation spread of 44\degree . Both can be scaled down to 14\degree\ at a different frequency.

\paragraph{Applying K-Factor, Delay Spread and Angle Spreads}
The initial delays $\tilde{\tau}_l$ and powers $\tilde{P}_{l,f}$ are chosen such that the \ac{DS} can have values between 0.15 and 0.85~seconds. The cluster angles $\tilde{\phi}_l^d$, $\tilde{\phi}_l^a$, $\tilde{\theta}_l^d$, $\tilde{\theta}_l^a$ are initialized such that the initial \acp{AS} can have values between 14\degree\ and 42\degree . Hence, the initial cluster powers already take the frequency-dependent differences for the \ac{DS} and \acp{AS} into account. Next, the absolute values of the \ac{KF}, the \ac{DS} and the four \acp{AS} are applied.

The \ac{KF} is the ratio of the power of the direct path to the sum-power of all other paths. It is applied by scaling the power of the first path and normalizing the \ac{PDP} to unit power. This is done independently for each frequency.
\begin{equation}\label{eq:path_powers_final_n}
    \tilde{P}_{1,f} = \text{KF}_f \cdot \sum_{l=2}^L \tilde{P}_{l,f} \qquad P_{l,f} = \tilde{P}_{l,f} / \sum_{l=1}^L \tilde{P}_{l,f}
\end{equation}

The \ac{DS} measures how the multipath power is spread out over time. Given the cluster-powers $P_{l,f}$ from \eqref{eq:path_powers_final_n} and the initial delays $\tilde{\tau}_{l}$ from \eqref{eq:initial_delays_n}, the initial \ac{DS} is calculated as
\begin{equation}\label{eq:ds_est}
   \widetilde{\text{DS}}_f = \sqrt{ \frac{1}{P_f} \cdot \sum_{l=1}^{L} P_{l,f} \cdot \left(\tilde{\tau}_{l}\right)^2 -     \left(\frac{1}{P_f}\cdot\sum_{l=1}^{L} P_{l,f} \cdot \tilde{\tau}_{l}\right)^2 }\text{,}
\end{equation}
where $P_f$ is the sum-power of all clusters at frequency $f$. The values of $\widetilde{\text{DS}}_f$ are frequency-dependent due to the scaling of the path-powers in \eqref{eq:multi_freq_cluster_powers}, but do not contain the correct \ac{DS} values from the \ac{LSF} model. Hence, the delays need to be scaled such that the correct \ac{DS} can be calculated from the generated path-delays and path-powers. This is done by
\begin{equation}
    \tau_l = \tilde{\tau}_{l} \cdot \frac{1}{F} \cdot \sum_{f=1}^F \frac{\text{DS}_f}{\widetilde{\text{DS}}_f}\text{.}
\end{equation}

The \ac{AS} measures how the multipath power is spread out in the spatial domain. The \ac{AS} is ambiguous since the angles are distributed on a sphere and the resulting value depends on the reference angle, \emph{i.e.}, the definition of where 0\degree\ is. A linear shift of the angles $\phi_{l} + \Delta_\phi$ leads to the \ac{AS} being a function of $\Delta_\phi$. We therefore normalize the angles such that the combined \acF{PAS} of all paths and sub-paths points to $\theta = \phi = 0$. The \ac{AS} is calculated by
\begin{eqnarray}
 \label{eq:mean_angle_est}
    \Delta_{\phi_f} &=& \arg\left(  \sum_{l=1}^L  \exp\left\{ j\tilde{\phi}_{l} \right\} \cdot P_{l,f} \right)\text{,}\\
 \label{eq:shifted_angles_est}
    \hat{\phi}_{l,f} &=& \arg \exp\left\{ j \left(\tilde{\phi}_{l} - \Delta_{\phi_f}\right) \right\}
\end{eqnarray}
\begin{equation} \label{eq:angular spread_est}
  \widetilde{\text{AS}}_f = \sqrt{ \frac{1}{P_f} \cdot \sum_{l=1}^L P_{l,f} \cdot \left(\hat{\phi}_{l,f}\right)^2 -
  \left( \frac{1}{P_f} \cdot \sum_{l=1}^L P_{l,f} \cdot \hat{\phi}_{l,f} \right)^2  }\text{.}
\end{equation}
With $\text{AS}_f$ being the initial \ac{AS} from the \ac{LSF} model, the initial angles $\tilde{\phi}_{l}$ are scaled to
\begin{multline}
    \phi_l = \arg \exp \left( j\cdot \tilde{\phi}_{l} \cdot s\right)\text{,} \quad s = \frac{1}{F} \cdot \sum_{f=1}^F \frac{\text{AS}_f}{\widetilde{\text{AS}}_f}\text{,} \quad \\
s < \left\{
      \begin{array}{ll}
        3.0, & \hbox{for azimuth angles;} \\
        1.5, & \hbox{for elevation angles.}
      \end{array}
    \right.
\end{multline}
The $\arg \exp$ function wraps the angles around the unit circle. The scaling coefficient $s$ is limited to a maximum value of 3 for the scaling of the azimuth angles and 1.5 for the elevation angles. This is motivated by the distribution of the initial angles in \eqref{eq:initial_angles_90}. More power is assigned to the angles having values close to 0 than to those having values close to $\pm \frac{\pi}{2}$. For this reason, $s=3$ achieves the maximum azimuth \ac{AS} of around 80\degree\ and $s = 1.5$ achieves the maximum elevation \ac{AS} of around 45\degree . Larger values of $s$ tend to decrease the \ac{AS} again due to the wrapping around the unit circle.

\paragraph{LOS angles} The last step is to apply the direction of the \ac{LOS} path. The initial values of the \ac{LOS} angles $\tilde{\phi}_1^d$, $\tilde{\phi}_1^a$, $\tilde{\theta}_1^d$, and $\tilde{\theta}_1^a$ were set to 0. However, the correct angles need to take the positions of the \ac{TX} and the \ac{RX} into account. The \ac{LOS} angles are
\begin{eqnarray}
  \phi_1^d &=& \arctan_2\left\{ y_r - y_t , x_r - x_t \right\}, \\
  \phi_1^a &=& \phi_1^d + \pi \\
  \theta_1^d &=& \arctan_2\left\{ z_r - z_t, d_{2d} \right\}, \\
  \theta_1^a &=& -\theta_1^d, \\
  d_{2d} &=& \sqrt{(x_r - x_t)^2 + (y_r - y_t)^2}
\end{eqnarray}
where $\arctan_2(y,x)$ is the four quadrant inverse tangent of the elements $y$ and $x$ having values between $-\pi$ and $\pi$. Those angles are applied by two \ac{3-D} rotations in Cartesian coordinates, one for the \ac{TX} and one for the \ac{RX}. The operations are identical. The NLOS departure and arrival angles from the previous calculation are converted to Cartesian coordinates by
\begin{equation}
    \mathbf{c}_l = \left(
                     \begin{array}{c}
                       \cos \theta_l \cdot \cos \phi_l \\
                       \cos \theta_l \cdot \sin \phi_l  \\
                       \sin \theta_l \\
                     \end{array}
                   \right)\text{.}
\end{equation}
Then, a rotation matrix is constructed from the \ac{LOS} angles. This matrix is a combined rotation around the $y$-axis followed by a rotation around the $z$-axis in Cartesian coordinates. It is applied by
\begin{equation}\label{eq:spherial_los_rotation}
    \mathbf{\hat{c}}_l =
    \left(
      \begin{array}{ccc}
        \cos \theta_1 \cdot \cos \phi_1 & -\sin \phi_1 & -\sin \theta_1 \cdot \cos \phi_1  \\
        \cos \theta_1 \cdot \sin \phi_1 &  \cos \phi_1 & -\sin \theta_1 \cdot \sin \phi_1 \\
        \sin \theta_1  & 0 & \cos \theta_1 \\
      \end{array}
    \right) \cdot \mathbf{c}_l\text{.}
\end{equation}
The final angles are then obtained by converting $\mathbf{\hat{c}}_l$ back to spherical coordinates.
\begin{eqnarray}
  \phi_l   &=& \arctan_2\left\{ \hat{c}_{l,y}, \hat{c}_{l,x} \right\} \\
  \theta_l &=& \arctan_2\left\{ \hat{c}_{l,z}, \sqrt{\hat{c}_{l,x}^2+\hat{c}_{l,y}^2} \right\}
\end{eqnarray}

\section{Numeric Evaluations and Discussion}\label{sec:ss_model_numeric}

The aim of this section is to show how well the propagation parameters from 3GPP-NR model can be mapped to delays, angles and powers of individual \acp{MPC} in a wireless channel. For the evaluation of the new \ac{SSF} model we used QuaDRiGa version 2.2 which implements both, the 3GPP-\ac{NR} baseline model and the new \ac{SSF} model presented in this paper.

\begin{figure}[h]
    \centering
    \footnotesize
    Delay Spread (DS)
    \includegraphics[width=89mm]{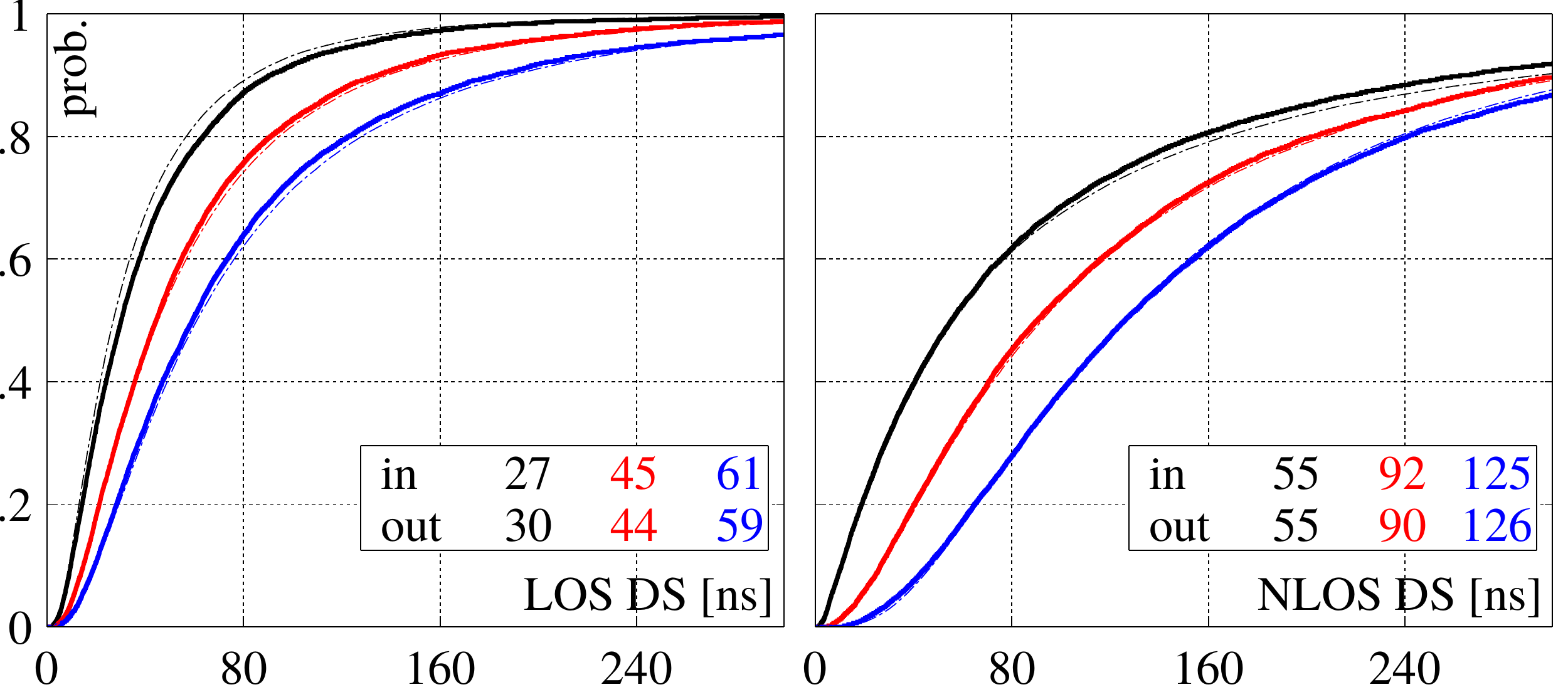}
    Azimuth of Departure Angular Spread (ASD)
    \includegraphics[width=89mm]{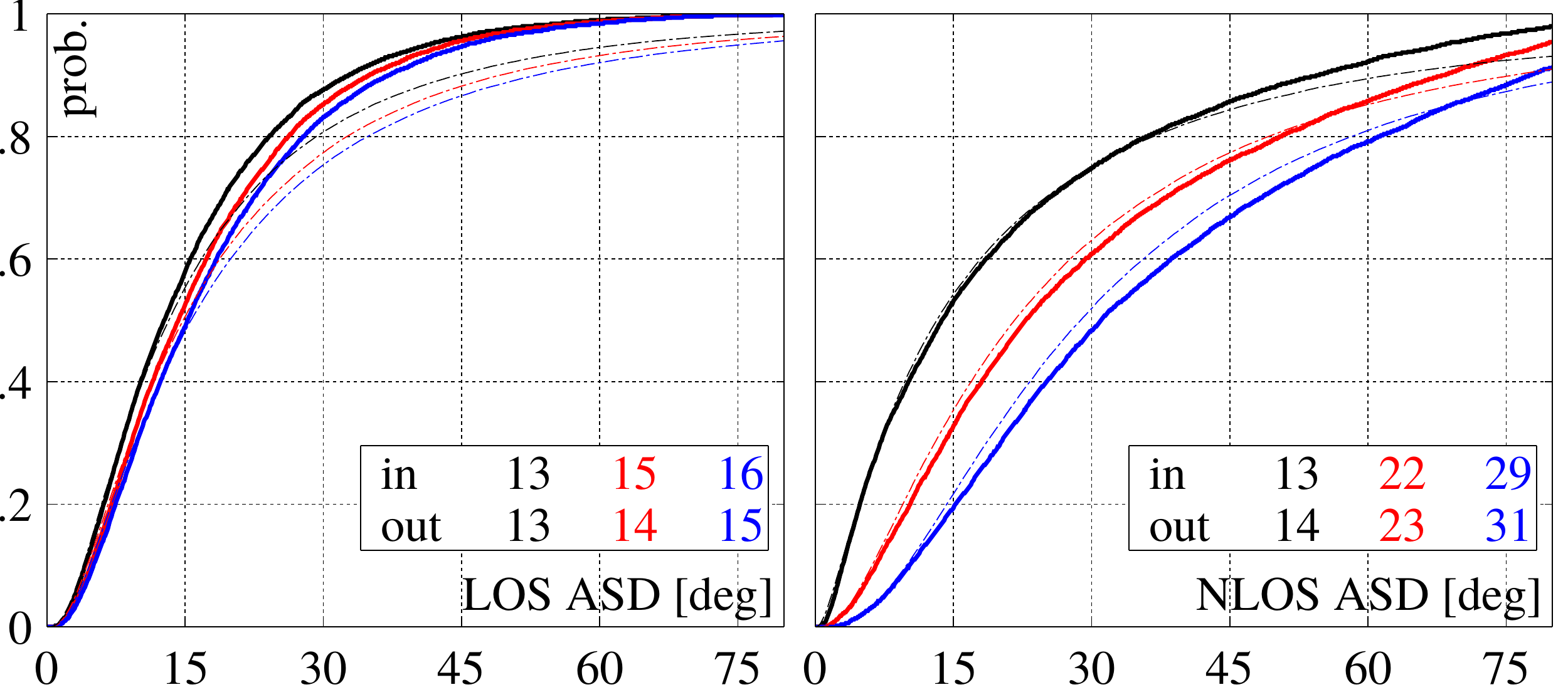}
    Azimuth of Arrival Angular Spread (ASA)
    \includegraphics[width=89mm]{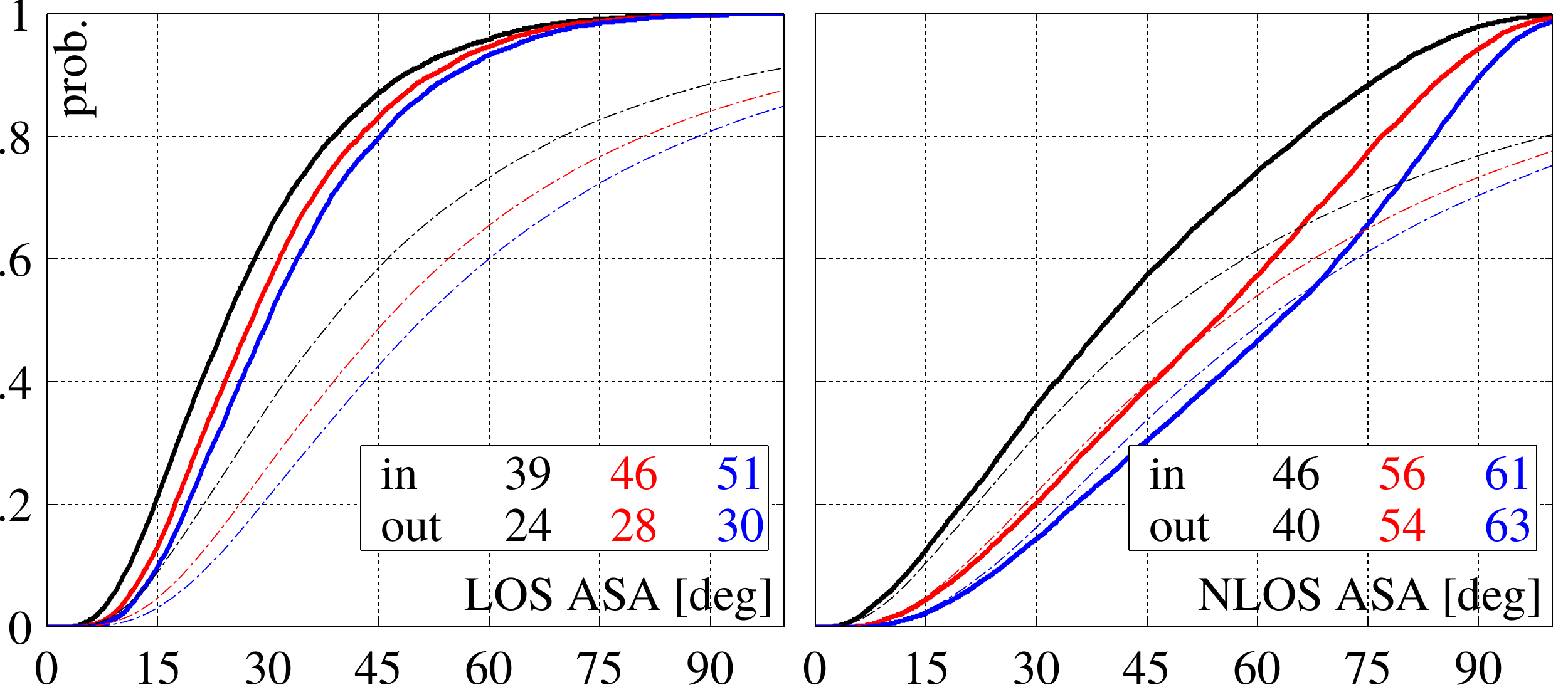}
    Elevation of Departure Angular Spread (ESD)
    \includegraphics[width=89mm]{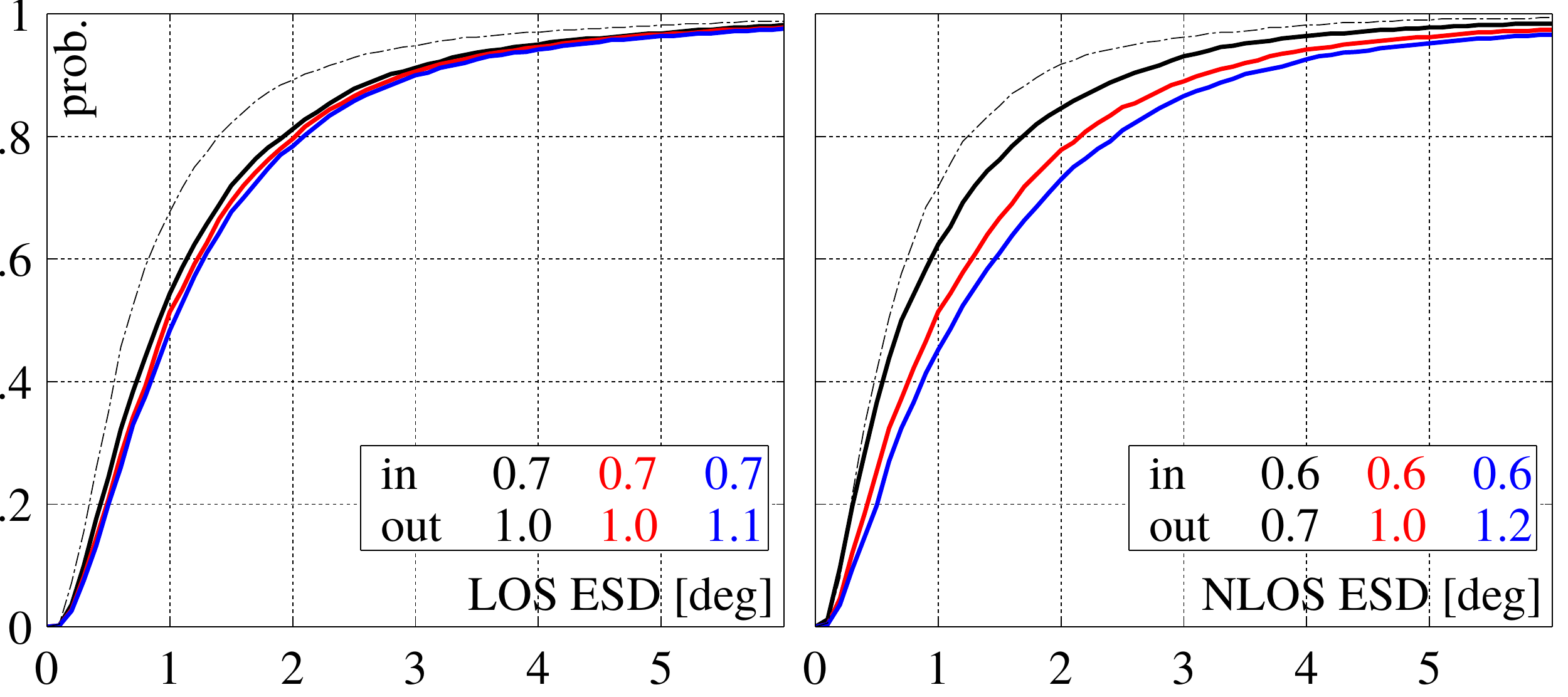}
    Elevation of Arrival Angular Spread (ESA)
    \includegraphics[width=89mm]{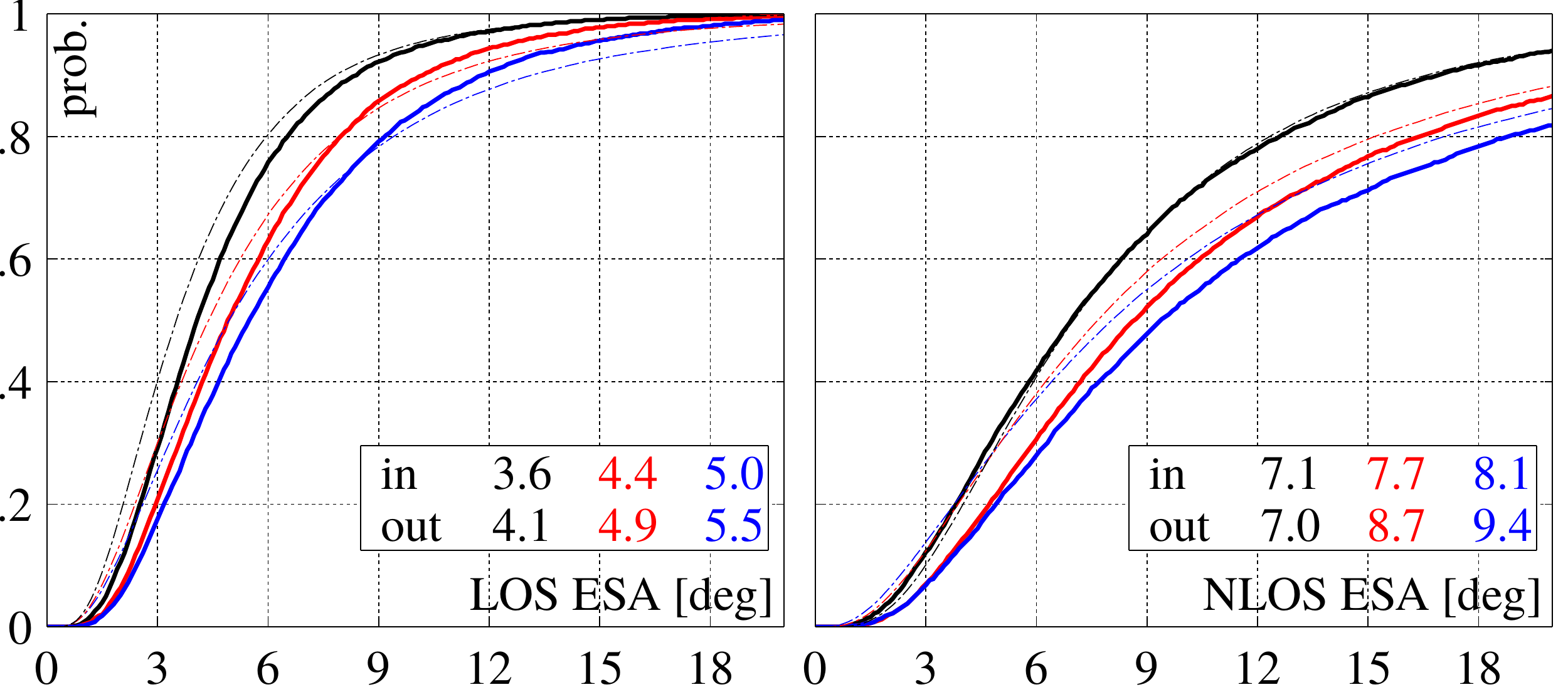}
    \vspace{-\baselineskip}
    \caption{Simulation Results for the 3GPP-NR UMi Scenario}
    \label{fig:ssf_scaling}
\end{figure}

The new \ac{SSF} model was evaluated for the \ac{UMi} scenario. The model parameters for \ac{LOS} and \ac{NLOS} propagation conditions are given by \cite{3gpp_tr_38901_v1500}, table 7.5-6. Evaluations were done for three different carrier frequencies: 1~GHz, 6~GHz, and 60~GHz. A single \ac{BS} was placed at a height of 10~m and 500~\acp{MT} were randomly placed within a 200~m radius around the \ac{BS}. The \ac{MT} height was set to 1.5~m. Both, the \ac{BS} and the \acp{MT} used isotropic antennas to remove the influence of the antenna patterns from the results. The evaluations were done as follows:
\begin{enumerate}
  \item Each \ac{MT} gets assigned a specific value of the \ac{DS}, the four \acp{AS} and the \ac{KF} as described in \cite{3gpp_tr_38901_v1500}, page 31, step 4. These values are the input to the new \ac{SSF} model.
  \item The input parameters get mapped to delays, angles and path-powers as described in Section~\ref{sec:ss_model} of this paper.
  \item The delay and angular spreads are calculated from the generated \acp{MPC} using \eqref{eq:ds_est} and \eqref{eq:angular spread_est}.
\end{enumerate}
As a result, we obtain the distribution of the input values from step 1 and the distribution of the output values from step 3. Ideally, these distributions are identical, i.e., the \ac{SSF} model maps \ac{LSF} parameters exactly. Results are shown in Fig.~\ref{fig:ssf_scaling}. The figure consists of 10 plots, each showing 6 empiric \acp{CDF}. The black curves show the results for 60~GHz, the red curves for 6~GHz and the blue curves for 1~GHz. Plots on the left-hand side are for \ac{LOS} propagation and plots on the right-hand side are for \ac{NLOS} propagation. The thin, dashed curves were obtained from the input values, i.e., the distributions given by \cite{3gpp_tr_38901_v1500}, table 7.5-6. The thick, solid curves were obtained from the output of the \ac{SSF} model. In addition, median values for the input and output are listed in the bottom right of each figure.

Except for the \ac{ESD}, all parameters are frequency-dependent where the values decrease with increasing frequency, i.e., the \ac{DS} is shorter at 60~GHz compared to 1~GHz. A generally good match between input and output of the new \ac{SSF} model can be achieved for most of the parameters. The \ac{DS} can be precisely mapped to path-delays and path-powers with ns-accuracy despite the fact that identical delay are used for all frequencies. The same holds true for the \ac{ASD} and the \ac{ESA} which is accurate within 1\degree . However, there are some significant offsets for the \ac{ASA} and the \ac{ESD}.

\begin{figure}[b]
    \centering
    \includegraphics[width=75mm]{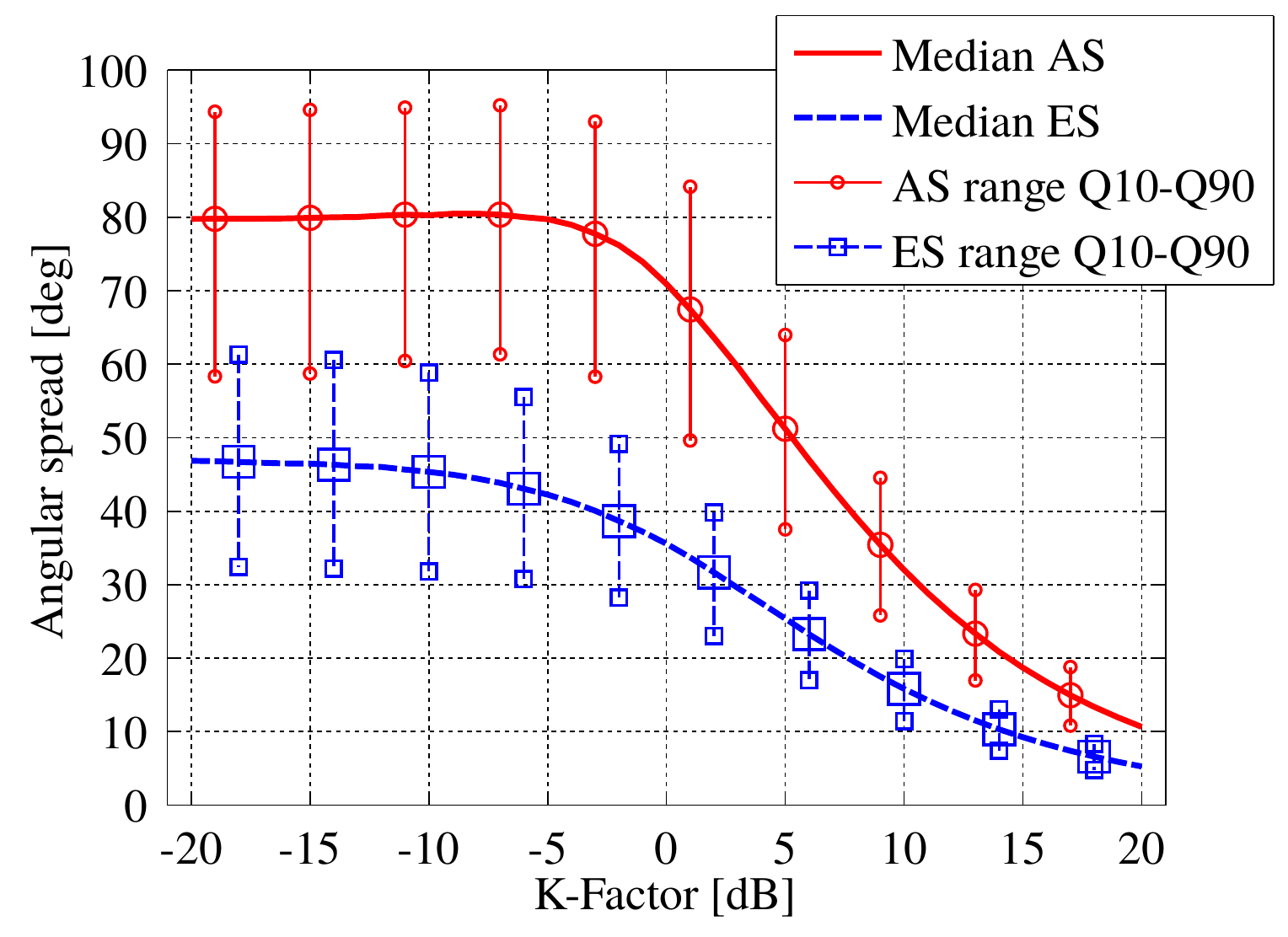}
    \vspace{-0.5\baselineskip}
    \caption{Maximal Angular Spread vs.\ K-Factor}
    \label{fig:max_angular_spread}
\end{figure}

Mapping the \acp{AS} of the \ac{LSF} model to scattering clusters fails when the azimuth angles have values outside the $-\pi$ to $\pi$ range or elevation angles exceed the $-\frac{\pi}{2}$ to $\frac{\pi}{2}$ range. This fact was also acknowledged by the 3GPP-NR model \cite{3gpp_tr_38901_v1500} where the azimuth \ac{AS} is capped at 104\degree\ and the elevation \ac{AS} is capped at 52\degree . However, this does not consider the influence of the \ac{KF} which limits the \ac{AS} even further when more power is allocated to the \ac{LOS} path. For example, with a \ac{KF} of 10 dB the maximum azimuth spread is 57\degree , provided that all \ac{NLOS} paths arrive from the opposite direction. Fig.~\ref{fig:max_angular_spread} shows the maximum \ac{AS} as a function of the \ac{KF} for our proposed model. The requested \ac{AS} was set to 100\degree . Those values cannot be achieved by the \ac{SSF} model. However, for \ac{NLOS} propagation, the achievable azimuth spread converges to values around 80\degree , and the elevation spread is around 45\degree . If the requested \ac{AS} is larger than the maximum \ac{AS}, the \ac{SSF} model adjusts the angles in a way that the \ac{AS} at the output of the model is close to the maximum \ac{AS}.

This \ac{KF}-dependency of \ac{AS} becomes a limiting factor for the \ac{ASA} in Fig.~\ref{fig:ssf_scaling}. For \ac{NLOS} propagation, the output curves start to diverge at around 60\degree\ and almost none of the output values exceed 95\degree . This agrees well with the results in Fig.~\ref{fig:max_angular_spread}. In the \ac{LOS} case, the 3GPP-NR model defines an average \ac{KF} of 9~dB with 5~dB \ac{STD} and an \ac{ASA} of around 50\degree . This is not achievable even under ideal circumstances. Still, the new \ac{SSF} model produces \ac{ASA} of around 30\degree\ as predicted by Fig.~\ref{fig:max_angular_spread}.

Lastly, there are some offsets for the \ac{ESD}, which is the only parameter which is not frequency-dependent. The 3GPP-NR model requires very low values around 0.6\degree\ which depend on the distance between \ac{BS} and \ac{MT}. However, the new \ac{SSF} almost doubles these values and introduces some frequency-dependency for the NLOS scenario. This comes from the application of the LOS angles in Sec.~\ref{sec:ss_model}-d. There is some mixing of the azimuth and elevation characteristics due to the spherical rotations introduced by  \eqref{eq:spherial_los_rotation}. However, this does not change the capacity of the wireless link since the overall \ac{AS} does not change.

\section{Conclusions}

The proposed modifications to the 3GPP-NR \acF{SSF} model enable the simultaneous simulation of wireless channels including spatial consistency, simultaneous mobility of the \acX{TX} and \acX{RX}, and the simultaneous simulation of multiple frequencies. This was not possible with the existing 3GPP-NR model which includes spatial consistency and multi-frequency simulations only as optional features and no \ac{D2D} modeling at all. However, these features are essential for many emerging wireless technologies in the 5G era. Our proposed \ac{SSF} model includes all these features and seamlessly integrates with the remaining \ac{NR} model. This has been demonstrated by comparing the output of the new \ac{SSF} model with the UMi \ac{LSF} parameter distributions provided by 3GPP. An open-source implementation of our model is provided to the community within \ac{QuaDRiGa} version 2.2 \cite{quadriga_www}.

\section*{Acknowledgement}

\small{The authors thank the Celtic Office and national funding authorities BMBF in Germany, Business Finland, and MINETAD in Spain for supporting this research and development through the ReICOvAir project. The research leading to these results has received funding from the European Union H2020 5GPPP under grant n.\ 815323 and supported by the Institute for Information \& communications Technology Promotion (IITP) grant funded by the Korea government (MSIT) (No.2018-0-00175, 5G AgiLe and fLexible integration of SaTellite And cellulaR).}

\bibliographystyle{IEEEtran}
\bibliography{ssf}

\end{document}

%% file: acronyms.tex
\newglossarystyle{acro}{
    \glossarystyle{super}

}
\newcommand{\acro}[2]{\newacronym{#1}{#1}{#2}}      
\newcommand{\acrol}[3]{\newacronym{#1}{#2}{#3}}     

\newcommand{\acrop}[4]{\newacronym[plural=#3,firstplural=#4 (#3)]{#1}{#1}{#2}}

\newcommand{\acF}[1]{\acf{#1}\glsunset{#1}}

\newcommand{\acX}[1]{\acrlong{#1}}

\acro{3GPP}{3rd generation partnership project}
\acrol{1-D}{\mbox{1-D}}{one-dimensional}
\acrol{2-D}{\mbox{2-D}}{two-dimensional}
\acrol{3-D}{\mbox{3-D}}{three-dimensional}
\acrol{4-D}{\mbox{4-D}}{four-dimensional}
\acrol{6-D}{\mbox{6-D}}{six-dimensional}
\acro{3G}{third generation}
\acro{4G}{fourth generation}
\acro{5G}{fifth generation}

\acrop{AoA}{azimuth angle of arrival}{AoAs}{azimuth angles of arrival}
\acrop{AoD}{azimuth angle of departure}{AoDs}{azimuth angles of departure}
\acro{ASA}{azimuth spread of arrival}
\acro{AS}{angular spread}
\acro{ASD}{azimuth spread of departure}
\acro{AWGN}{additive white Gaussian noise}
\acro{AGC}{adaptive gain control}
\acro{ACF}{autocorrelation function}
\acro{ASE}{average squared error}

\acro{BD}{block diagonalization}
\acro{BS}{base station}
\acro{BC}{broadcast channel}
\acro{BPSK}{binary phase-shift keying}
\acro{BP}{break point}

\acro{CDF}{cumulative distribution function}
\acro{CIR}{channel impulse response}
\acro{CoMP}{coordinated multi-point}
\acro{CoF}{cost of feedback}
\acro{CRONOS}{cellular radio network simulator}
\acro{CSI}{channel state information}
\acro{COST}{European Cooperation in Science and Technology}
\acro{CPS}{cyber physical system}

\acro{DS}{delay spread}
\acro{DD}{Doppler-delay}
\acro{DFT}{discrete Fourier transform}
\acro{DPC}{dirty-paper coding}
\acro{D2D}{device-to-device}

\acro{ESA}{elevation spread of arrival}
\acro{ESD}{elevation spread of departure}
\acrop{EoA}{elevation angle of arrival}{EoAs}{elevation angles of arrival}
\acrop{EoD}{elevation angle of departure}{EoDs}{elevation angles of departure}

\acro{FBR}{feedback rate}
\acro{FBS}{first-bounce scatterer}
\acro{FI}{feedback interval}
\acro{FR}{frequency response}
\acro{FR1}{frequency reuse 1}
\acro{FR4}{frequency reuse 4}
\acro{FIR}{finite impulse response}
\acro{FFT}{fast Fourier transform}
\acro{FWHM}{full width at half maximum}
\acro{FDD}{frequency-division duplexing}

\acro{GCS}{global coordinate system}
\acro{GF}{geometry factor}
\acro{GR}{ground reflection}
\acro{GoC}{gain of CoMP}
\acro{GoP}{gain of prediction}
\acro{GPS}{global positioning system}
\acro{GSCM}{geometry-based stochastic channel model}

\acro{HHI}{Heinrich Hertz Institute}
\acro{HPBW}{half-power beamwidth}

\acro{ISD}{inter site distance}
\acrol{iid}{i.i.d.}{independent and identically distributed}
\acro{IDFT}{inverse discrete Fourier transform}
\acro{IFFT}{inverse fast Fourier transform}
\acro{IR}{impulse response}
\acro{ITU}{International Telecommunication Union}
\acro{IF}{intermediate frequency}

\acro{JT}{joint transmission}
\acro{JCF}{joint correlation function}

\acro{KF}{Ricean K-factor}
\acro{KPI}{key performance indicator}

\acro{LBS}{last-bounce scatterer}
\acro{LHCP}{left hand circular polarized}
\acro{LOS}{line of sight}
\acro{LSP}{large-scale parameter}
\acro{LTE}{long term evolution}
\acro{LSAS}{large-scale antenna systems}
\acro{LSF}{large-scale fading}

\acro{MET}{multi-user eigenmode transmission}
\acro{MIMO}{multiple-input multiple-output}
\acro{MISO}{multiple-input single-output}
\acro{MMSE}{minimum mean square error}
\acro{MPC}{multipath component}
\acro{MS}{mobile station}
\acro{MSE}{mean square error}
\acro{MT}{mobile terminal}
\acro{MU-MIMO}{multi-user MIMO}
\acro{MIMOSA}{MIMO over satellite}
\acro{MAC}{multiple-access channel}
\acro{MRT}{maximum ratio transmission}

\acro{NLOS}{non-line of sight}
\acro{NR}{new radio}

\acro{OFDM}{orthogonal frequency division multiplexing}
\acro{O2I}{outdoor-to-indoor}
\acro{ODA}{omni directional array}

\acro{PDP}{power delay profile}
\acro{PDF}{probability density function}
\acro{PG}{path gain}
\acro{PL}{path loss}
\acro{PUCA}{polarized uniform circular array}
\acro{PAS}{power-angular spectrum}
\acro{P2P}{peer-to-peer}
\acro{PSCS}{Propsound channel sounder}

\acro{QAM}{quadrature amplitude modulation}
\acro{QuaDRiGa}{quasi deterministic radio channel generator}
\acro{QRDRLS}{QR-decomposition recursive least squares}
\acro{QoS}{quality of service}

\acro{RB}{resource block}
\acro{RHCP}{right hand circular polarized}
\acro{RLS}{recursive least-squares}
\acro{RLE}{run-length encoding}
\acro{RX}{receiver}
\acro{RMS}{root mean square}
\acro{RAN}{radio access network}
\acro{RSRP}{reference signal received power}
\acro{RF}{radio frequency}

\acro{SAGE}{space-alternating generalized expectation-maximization}
\acro{SCM}{spatial channel model}
\acro{SF}{shadow fading}
\acro{SV}{singular value}
\acro{SVD}{singular value decomposition}
\acro{SNR}{signal to noise ratio}
\acro{SIR}{signal to interference ratio}
\acro{SINR}{signal to interference and noise ratio}
\acro{SISO}{single input single output}
\acro{SIMO}{single input multiple output}
\acro{STD}{standard deviation}
\acro{SSG}{state sequence generator}
\acro{SSF}{small-scale-fading}
\acro{SOS}{sum-of-sinusoids}

\acro{TP}{throughput}
\acro{TUB}{Technische Universit\"{a}t Berlin}
\acro{TX}{transmitter}

\acro{ULA}{uniform linear array}
\acro{UML}{unified modeling language}
\acro{UMi}{urban-microcell}
\acro{UMa}{urban-macrocell}
\acro{UAV}{unmanned aerial vehicle}

\acro{VR}{visibility region}

\acro{WINNER}{Wireless World Initiative for New Radio}
\acro{WSSUS}{wide sense stationary uncorrelated scattering}
\acro{WSS}{wide-sense stationary}
\acro{WGS}{world geodetic system}

\acro{XPD}{cross-polarization discrimination}
\acro{XPR}{cross polarization ratio}

\acro{ZF}{zero-forcing}
\acro{ZoA}{zenith angle of arrival}
\acro{ZoD}{zenith angle of departure}
\acro{ZSA}{zenith angle spread of arrival}
\acro{ZSD}{zenith angle spread of departure}
\acro{ZCR}{zero-crossing rate}